\documentclass{aa}  

\usepackage{graphicx}
\usepackage{txfonts}
\begin{document}

   \title{Revisiting the tension between fast bars \& the $\Lambda$CDM paradigm}

   \author{F. Fragkoudi
          \inst{1,2}\thanks{\email{francesca.fragkoudi@eso.org}}
          \and
          R. J. J. Grand\inst{2}\fnmsep
          \and
          R. Pakmor$^{2}$
          \and
          V. Springel$^{2}$ 
          \and
          S. D. M. White$^{2}$ 
          \and
          F. Marinacci$^{3}$ 
          \and
          F. A. Gomez$^{4,5}$
          \and 
          \\
          J. F. Navarro$^{6}$
          }

   \institute{European Southern Observatory,
              Karl-Schwarzschild-Str. 2, 85748 Garching-bei-M\"unchen, Germany\\
              \email{francesca.fragkoudi@eso.org}
         \and
             Max-Planck-Institut f\"{u}r Astrophysik, Karl-Schwarzschild-Str. 1, 85748 Garching, Germany
        \and
            Department of Physics and Astronomy, University of Bologna, via Gobetti 93/2, I-40129 Bologna, Italy
         \and
            Instituto de Investigacion Multidisciplinar en Ciencia y Tecnolog\'{i}a, Universidad de La Serena, Raul Bitr\'{a}n 1305, La Serena, Chile
        \and
            Departamento de Astronom\'ia, Universidad de La Serena, Av. Juan Cisternas 1200 Norte, La Serena, Chile
        \and
            Department of Physics and Astronomy, University of Victoria, Victoria, BC, V8P 5C2, Canada
             }

   \date{Received xx xx; Accepted xx xx}

 
  \abstract
{The pattern speed with which galactic bars rotate is intimately linked to the amount of dark matter in the inner regions of their host galaxies. In particular, dark matter haloes act to slow down bars via torques exerted through dynamical friction. 
Observational studies of barred galaxies tend to find that bars rotate fast, while hydrodynamical cosmological simulations of galaxy formation and evolution in the $\Lambda$CDM framework have previously found that bars slow down excessively. This has led to a growing tension between fast bars and the $\Lambda$CDM cosmological paradigm.
In this study we revisit this issue, using the Auriga suite of high resolution, magneto-hydrodynamical cosmological zoom-in simulations of galaxy formation and evolution in the $\Lambda$CDM framework, finding that bars remain fast down to $z=0$. In Auriga, bars form in galaxies that have higher stellar-to-dark matter ratios and are more baryon-dominated than in previous cosmological simulations; this suggests that in order for bars to remain fast, massive spiral galaxies must lie above the commonly used abundance matching relation.
While this reduces the aforementioned tension between the rotation speed of bars and $\Lambda$CDM, it accentuates the recently reported discrepancy between the dynamically inferred stellar-to-dark matter ratios of massive spirals and those inferred from abundance matching. Our results highlight the potential of using bar dynamics to constrain models of galaxy formation and evolution.}


   \maketitle
%

\section{Introduction}

   Bars are common structures in spiral galaxies in the local Universe \citep{Eskridgeetal2000,Menendezetal2007} and are able to redistribute angular momentum \citep{LyndenBellKalnajs1972} from the inner regions of the disc to the outer disc and dark matter halo. The distribution function of the disc and halo determine the amount of material which is available to `emit' and `absorb' this angular momentum \citep{Athanassoula2003}. 
The density distribution of dark matter in the inner regions of haloes is therefore one of the main factors driving the formation and evolution of bars themselves \citep{DebattistaSellwood2000}, while other important factors include the velocity dispersion of the disc and halo, the central mass concentration and the gas fraction in the disc \citep{OstrikerPeebles1973,Combesetal1981,Athanassoulaetal2013}. 

The inner regions of stellar discs lose angular momentum via the bar, by trapping stars onto elongated orbits -- and therefore growing in mass and length over time \citep{Athanassoula2003} -- and through dynamical friction with the dark matter halo \citep{TremaineWeinberg1984DF,Weinberg1985}.
Dynamical friction can act to slow down the angular rotation speed of the bar, commonly referred to as the bar pattern speed, $\Omega_{\rm p}$, without increasing the length of the bar, $R_{\rm bar}$ \citep{DebattistaSellwood2000}. When bars slow down, the corotation radius, $R_{\rm CR}$ -- the radius at which stars on near circular orbits move around the galaxy with the same angular frequency as the bar -- increases. As a result, the corotation radius normalised by the bar length, $\mathcal{R}=R_{\rm CR}/R_{\rm bar}$, can shed light on how much dynamical friction the halo exerts on the bar, and therefore on the amount of dark matter in the inner galaxy. As is common in the literature, in what follows we will refer to bars with $\mathcal{R}<1.4$ and $\mathcal{R}>1.4$ as ``fast'' and ``slow'' respectively \citep{DebattistaSellwood2000}.

Numerous observational efforts have been made to measure the pattern speed of bars, finding that bars tend to rotate fast, i.e. have $\mathcal{R}<$1.4  \citep{Corsini2011,Aguerrietal2015,Guoetal2019}. 
This suggests that dark matter haloes do not exert much dynamical friction on bars, and are therefore likely subdominant in the central regions of galaxies. However, uncertainties in obtaining the mass-to-light ratio (M/L) of stellar discs make determining the baryon-to-dark matter ratio in the inner regions of galaxies difficult. Dynamical studies of massive spiral galaxies in the local Universe tend to find that these are baryon-dominated in their central regions \citep{Weiner2001,Kranzetal2003,BovyRix2013,Fragkoudietal2017a}. This is also found by studies which use the less uncertain M/L obtained from stellar population models in the near-infrared \citep{Lellietal2016}. On the other hand, studies such as the DiskMass Survey point to contradictory results, suggesting that dark matter haloes dominate in the central regions of spiral galaxies \citep{Bershadyetal2011}. 
Therefore, whether or not spiral galaxies are baryon-dominated in their central regions, is still under debate. 

On the other hand, advances in numerical and physical implementations have led to drastic improvements in hydrodynamical cosmological simulations, which are now  able to routinely form spiral galaxies with extended discs \citep{Agertzetal2011,Vogelsbergeretal2014,Schayeetal2015,Grandetal2017}. However, the relative contribution of baryonic and dark matter in the inner regions of galaxies in $\Lambda$CDM is an ongoing topic of debate, as baryonic processes play an important role in reshaping the inner profiles of dark matter haloes and in setting the baryon-to-dark matter ratio in the central regions \citep{PontzenGovernato2012,Lovelletal2018}. 
Studying the dynamics of barred galaxies in cosmological simulations provides a powerful tool for constraining the relative amount of baryons and dark matter in the inner regions.

There have been a handful of studies which have explored the properties of bars in the full $\Lambda$CDM cosmological context to date (e.g. \citealt{Kraljicetal2012,ScannapiecoAthanassoula2012,Zanaetal2018,RosasGuevaraetal2020,BlazquezCaleroetal2020}). However, few of these have explored their pattern speeds, and only two explore this in a large sample of galaxies \citep{Algorryetal2017,PeschkenLokas2019}, using the EAGLE \citep{Schayeetal2015} and Illustris \citep{Vogelsbergeretal2014} simulations. These studies found that by $z=0$, bars have slowed down excessively, making them incompatible with observations. This has resulted in a growing tension between $\Lambda$CDM cosmological simulations and fast bars, raising the question: \emph{are fast bars incompatible with the $\Lambda$CDM framework?}

Here we revisit the question of the slow-down of bars in the $\Lambda$CDM context, using the Auriga simulations, a suite of state-of-the-art high-resolution magneto-hydrodynamical cosmological zoom-in simulations of the formation of Milky Way-mass galaxies. In Section \ref{sec:ausims} we show that bars in Auriga remain fast until $z=0$, compatible with observations. In Section \ref{sec:4} we compare our results to previous findings in the literature, and discuss some of the differences in the simulations that can give rise to the different dynamical behaviour. In Section \ref{sec:concl} we discuss some of the implications of our findings and conclude.

\section{Fast-rotating bars in the Auriga Simulations}
\label{sec:ausims}

The Auriga simulations \citep{Grandetal2017} are a suite of 30 magneto-hydrodynamical cosmological zoom simulations of haloes with masses in the range of $1 \times 10^{12}-2 \times 10^{12}M_{\odot}$.
By $z=0$, the simulations form star-forming disc galaxies with flat rotation curves that reproduce a range of observed scaling relations such as the Tully-Fisher relation \citep{Grandetal2017} and the size-mass relation of HI gas discs \citep{Marinaccietal2017}. They also form structures such as bars and bulges which have properties compatible with those of observed galaxies \citep{BlazquezCaleroetal2020,Fragkoudietal2020,Gargiuloetal2019}. For this study our barred sample includes all the Auriga galaxies with bar strength $>0.2.$ at $z=0$, where bar strength is defined as the maximum amplitude of the $m=2$ Fourier mode of the surface density. We exclude the galaxies which are undergoing an interaction at $z=0$.  For more details on the sample of Auriga simulations and how the bar strength is derived, we refer the reader to Appendix \ref{sec:appAu} \& \ref{sec:ap1}.

\begin{figure}
\centering
\includegraphics[width=0.49\textwidth]{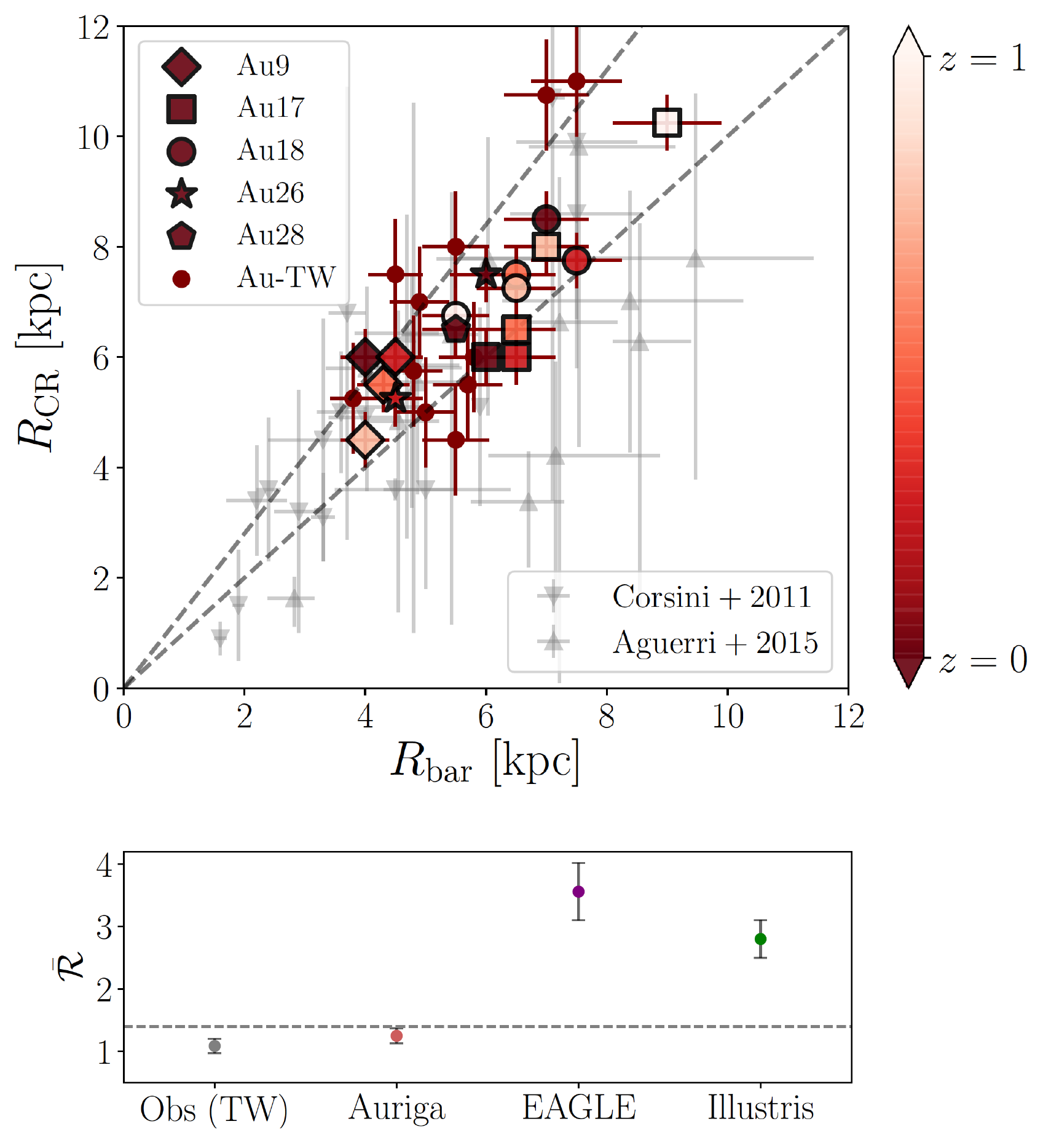}
\caption{{\textbf{Bar corotation radius vs bar length for Auriga galaxies:}} \emph{Top panel:} Corotation radius as a function of bar length for the barred Auriga galaxies, denoted by the symbols in the figure legend. The colour-coding corresponds to measurements at different redshifts, while for the galaxies for which we use the TW method (Au-TW) we estimate the pattern speed at $z=0$. These are compared to data from Corsini (2011) and Aguerri et al. (2015) (grey points); the error bars indicate the 1$\sigma$ uncertainties. The lower and upper dashed grey lines indicate $\mathcal{R}=1$ and $\mathcal{R}=1.4$ respectively. Bars in Auriga are fast across all redshifts. \emph{Bottom panel:} We show the mean $\mathcal{R}$ for the aforementioned observations, for barred galaxies in Auriga, and for the EAGLE and Illustris barred galaxies from Algorry et al. (2017) and Peschken et al. (2019) at $z=0$. The error bars indicate the 2$\sigma$ error on the mean. The dashed line indicates $\mathcal{R}=1.4$, below which bars are considered fast. Bars in the Auriga simulations are fast, compatible with observations, while bars in EAGLE and Illustris are excessively slow at $z=0$.} 
\label{fig:Omprbar}
\end{figure}

As mentioned above, while the absolute value of the pattern speed of bars can shed light on the exchange of angular momentum in the galaxy, the parameter used to determine the bar slow-down due to dynamical friction is $\mathcal{R}$ = $R_{\rm CR}$/$R_{\rm bar}$ due to the fact that dynamical friction can act to slow down the bar without a corresponding increase in its length. 
Calculating the length of bars is non-trivial as it oscillates due to the coupling of different structures in the disc, such as the bar with the spiral arms (e.g. \citealt{Petersenetal2019,Hilmietal2020}). For the estimate of $R_{\rm bar}$ we employ a method similar to that used in analysing observations, i.e. the ellipse fitting method in \cite{Erwin2005}. We tested however that our results on $\mathcal{R}$ do not change if we opt for a different bar length estimate (see Appendix \ref{sec:ap1} for details on the bar length calculation and various tests). The pattern speed in Auriga is calculated from the temporal evolution of the $m=2$ Fourier phase of the surface density, or via the Tremaine-Weinberg method (TW; \citealt{TremaineWeinberg1984}) for simulations without high cadence outputs (see Appendix \ref{sec:ap1} for more details).

We plot $R_{\rm CR}$ versus $R_{\rm bar}$ in the top panel of Figure \ref{fig:Omprbar} for the barred Auriga galaxies and compare these to observations \citep{Corsini2011,Aguerrietal2015}, in which the bar pattern speed is obtained using the TW method. 
The upper and lower solid lines in Figure \ref{fig:Omprbar} correspond to $\mathcal{R}$ = 1.4 and 1 respectively, which denote the regime in which bars are considered `fast'. For the five Auriga galaxies for which we have high cadence outputs (see Appendix \ref{sec:appAu}) we calculate the corotation radius and bar length at redshifts $z=0,0.25,0.5,0.75,1$ (provided the bar has already formed by the corresponding redshift), as denoted by the colour-coding of the symbols. We find that when bars in the Auriga galaxies are formed, they have $\mathcal{R}<1.4$, i.e. they are dynamically fast, and that they remain so throughout their evolution. We note that 25\% of the sample have $\mathcal{R}$ slightly higher than 1.4, but are compatible with 1.4 within the errors. We therefore find that the Auriga bars are on average fast at $z=0$, compatible with observations.

\section{Comparison with previous pattern speed estimations from cosmological simulations}
\label{sec:4}

In the bottom panel of Figure \ref{fig:Omprbar} we show the mean and 2$\sigma$ error on the mean of $\mathcal{R}$, for barred galaxies in the Auriga simulations, for the above-mentioned observations, as well as for the barred galaxies studied previously in $\Lambda$CDM cosmological simulations \citep{Algorryetal2017,PeschkenLokas2019}. As was shown by these latter works, bars in the EAGLE and Illustris simulations tend to have $\mathcal{R}>2.5$, in tension with observations. Bars in Auriga have $\mathcal{R} < 1.4$, thus demonstrating that bars can remain dynamically fast in cosmological simulations within the $\Lambda$CDM paradigm. This therefore lessens, to some extent, the previously reported tension between observed fast bars and $\Lambda$CDM. 

However a natural question to ask is `what gives rise to the different behaviour of $\mathcal{R}$ in the different simulations?'
As mentioned above, $\mathcal{R}$ depends on the amount of dynamical friction exerted by the halo on the disc, which in turn depends strongly on the amount of dark matter present in the central region of the halo, as compared to the disc.
The subgrid physics modelling employed in cosmological simulations can therefore play a critical role in determining the bar properties, and more crucially, its slowdown rate, as it sets the disc-to-halo mass ratio in the central regions of galaxies. In what follows we investigate the effect of numerical resolution, as well as the properties of the galaxies in the different simulations that might lead to differences in $\mathcal{R}$.  

\paragraph{Resolution:}

\begin{figure}
\centering
\includegraphics[width=0.49\textwidth]{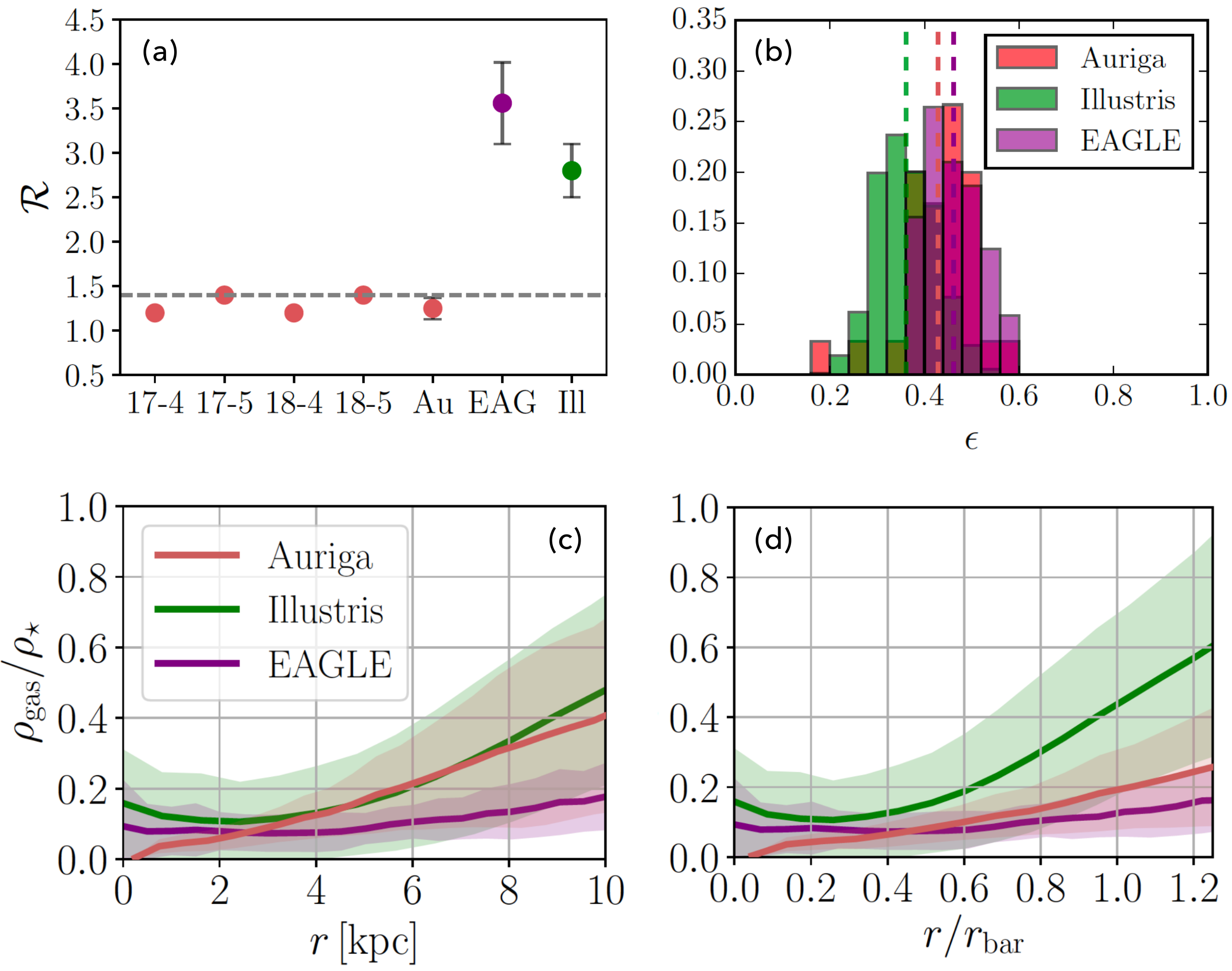}
\caption{{\bf{(a) Resolution test for $\mathcal{R}$:}} The value of $\mathcal{R}$ for the two lower resolution runs (Au17-5 and Au18-5) compared to the standard resolution runs (Au17-4 and Au18-4). For reference we also show the mean and 2$\sigma$ error of $\mathcal{R}$ for the whole bar sample of Auriga, EAGLE and Illustris. {\bf{ (b) Disc thickness:}} The flattening,  $\epsilon=1-c/a$, of disc galaxies in Auriga, Illustris and EAGLE. The average values of the distributions (vertical dashed lines) are 0.42, 0.36 and 0.46 respectively. {\bf{(c) Gas fraction vs radius:}} The average gas density over stellar density in the Auriga galaxies as compared to disc galaxies in EAGLE and Illustris. {\bf{(d) Gas fraction vs normalised radius:}} As in panel (c), with the radius normalised by the average bar length in each of the simulations.} 
\label{fig:restest}
\end{figure}

Low numerical resolution has been extensively discussed in the literature as a parameter which could affect the evolution of bars and the exchange of angular momentum at resonances, often with different studies reaching contradictory conclusions \citep{Weinberg1998,WeinbergKatz2002,ValenzuelaKlypin2003,Sellwood2006,Sellwood2008}.

 In order to test the effect of resolution in our models, and to explore whether it could be the primary reason for why previous studies using the EAGLE and Illustris cosmological simulations found large values of $\mathcal{R}$, we carry out a resolution test. We re-ran two of our Auriga haloes (Au17 and Au18) with 8$\times$ lower mass resolution, and 2$\times$ lower spatial resolution. This leads to a mass resolution that is closer to -- although still slightly higher than -- the EAGLE and Illustris resolution. In the low resolution Auriga runs the stellar and dark matter particles have a mass of 4$\times10^5M_{\odot}$ and 3.2$\times10^6M_{\odot}$ respectively, while in EAGLE (Illustris) the stellar mass is  1.8$\times10^6M_{\odot}$ (1.3$\times10^6M_{\odot}$) and the dark matter particle mass is 9.7$\times10^6M_{\odot}$ (6.3$\times10^6M_{\odot}$). If resolution is the major contributor to the high values of $\mathcal{R}$ in previous studies we expect to see a significant effect on $\mathcal{R}$ when decreasing the resolution to be close to that of EAGLE and Illustris.
 
 In Figure \ref{fig:restest}a we show the $\mathcal{R}$ values for the level 4 (high resolution) and level 5 (low resolution) runs for Au17 and Au18. We find that $\mathcal{R}$ has a slight increase of $\sim$15\% in the low resolution runs. However, the increase is within the 2$\sigma$ error of the mean $\mathcal{R}$ values for the high resolution barred galaxies explored in this study. Therefore, this small increase in $\mathcal{R}$ for the low resolution Auriga runs does not appear to be sufficient to explain the much higher values found in the EAGLE and Illustris simulations\footnote{It is also worth noting that high-resolution zoom simulations in the literature (e.g. \citealt{Zanaetal2018}), which study the pattern speed of bars also find $\mathcal{R}>1.4$ at $z=0$.}. These considerations suggest that resolution is likely not the main culprit for the high $\mathcal{R}$ values found in previous simulations such as EAGLE and Illustris.

\begin{figure*}
\centering
\includegraphics[width=0.97\textwidth]{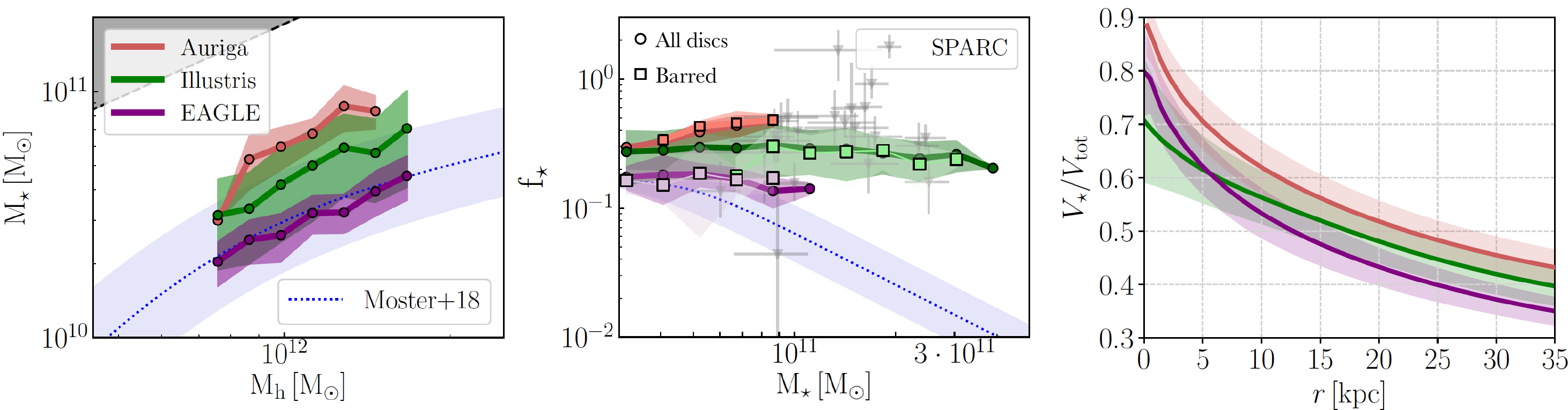}
\caption{{\bf{Global and local stellar-to-dark matter ratios in the Auriga galaxies as compared to EAGLE and Illustris:}} \emph{Left:} The mean (lines) and 1$\sigma$ (shaded region) values of the $M_{\star}$ vs $M_{\rm h}$ of Auriga, EAGLE and Illustris disc galaxies at $z=0$ compared to the abundance matching relation from Moster et al. (2018). 
\emph{Middle:} The efficiency with which gas is turned into stars, i.e. $f_{\star}=\frac{M_{\rm \star}}{f_b M_{\rm h}}$ as a function of stellar mass for the Auriga, EAGLE and Illustris galaxies, compared to the relation from abundance matching. The mean and 1$\sigma$ values for disc (barred) galaxies are shown with the circles/thick lines (squares/thin lines) and shaded region respectively.  For comparison we also show the high mass disc galaxies from the SPARC sample explored in Marasco et al. (2020), where the error bars denote 16th-84th percentile uncertainties.
\emph{Right:} The local baryon dominance, $V_{\star}/V_{\rm tot}$ as a function of radius, of Auriga, EAGLE and Illustris disc galaxies in the mass range $3\times 10^{10}<M_{\star}/M_{\odot}<10^{11}$.
} 
\label{fig:dmbarratio}
\end{figure*}

\paragraph{Disc thickness:}
The thickness of discs -- which is related to their vertical velocity dispersion -- has been reported in the literature as playing a role in the slow down of bars (\citealt{MisiriotisAthanassoula2000,Klypinetal2009}). While the exact mechanism for this is still unclear, it has been suggested that bars are slower in thicker discs (\citealt{Klypinetal2009}). In Figure \ref{fig:restest}b we investigate the differences in disc thickness between Auriga, Illustris and EAGLE, by exploring the flattening of their mass distributions, defined as $\epsilon = 1 - c/a$, where $c$ ($a$) corresponds to the smallest (largest) axis, i.e. $\epsilon=0$ corresponds to a spheroidal distribution. We obtain the axes of the ellipsoid by calculating the reduced inertia mass tensor of the galaxies (for more details see Appendix \ref{sec:dth}).  We find that the disc galaxies in Auriga, Illustris and EAGLE have similar distributions of flattening, with Illustris having a slightly lower average value (0.36) than Auriga (0.42) and EAGLE (0.46). While a slight difference in disc thickness might be contributing to the slow down of bars in Illustris, the flattening of discs in Auriga and EAGLE is very similar, and it is therefore unlikely that disc thickness alone can explain the bar slowdown.

\paragraph{Gas fraction:}

The fraction of gas within discs, and in particular within the bar radius, can play a role in both the formation and strength of bars. It has also been suggested that gas can play a role in maintaing fast bars \citep{Athanassoula2014}, although the topic has given rise to considerable debate (e.g. \citealt{Bournaudetal2005,Berentzenetal2007,Athanassoula2014,Sellwoodetal2014} and references therein). In the two lower panels of Figure \ref{fig:restest} we explore the gas content of disc galaxies in Auriga, EAGLE and Illustris. Auriga and Illustris have similar gas fractions within 10\,kpc, while EAGLE has lower gas fractions, which could possibly contribute to the slower bars in EAGLE as compared to Auriga. However, when the radius is normalised by the average bar length in each simulation, we find that, within the bar radius, Auriga galaxies have similar gas fractions to EAGLE, and lower gas fractions than Illustris. Therefore, while the gas fraction might contribute to the differences in pattern speed between the different simulations, it is likely not the main culprit for allowing Auriga to maintain fast bars.

\paragraph{Baryon-dominance}


We therefore turn our attention to the relative contribution of the stellar component compared to dark matter in the Auriga simulations. We explore this in Figure \ref{fig:dmbarratio} and compare Auriga to the EAGLE and Illustris cosmological simulations. In the left panel we plot the average and 1$\sigma$ relation for the stellar mass as a function of the dark matter halo mass ($M_{\rm \star}$ vs $M_{\rm h}$) for the Auriga galaxies (red), as well as for the Illustris and EAGLE discs (green and purple respectively). We compare these to the commonly used abundance matching relation (\citealt{Mosteretal2018}; dashed blue line), which is obtained by matching the observed stellar mass function to the halo mass function in cosmological simulations. Auriga galaxies are offset from the relation, and lie above both the EAGLE and Illustris discs, i.e. they are globally more baryon-dominated.

In the middle panel of Figure \ref{fig:dmbarratio} we plot the efficiency with which -- given a universal baryon-to-dark matter ratio $f_{\rm b}=\Omega_{\rm b}/\Omega_{\rm c}$ -- galaxies convert their gas into stars, i.e. $f_{\rm \star}=M_{\rm \star}/(f_{\rm b}M_{\rm h})$, as a function of stellar mass. This is shown for the Auriga, EAGLE and Illustris simulations, for both the entire disc sample (circles) and for the barred galaxies (squares). 
The Auriga galaxies are offset from the relation predicted by abundance matching -- by more than 2$\sigma$ at $M_{\star} = 9\times 10^{10}M_{\odot}$. They also lie above both the EAGLE and Illustris galaxies, and thus have higher global stellar-to-dark matter ratios for a given stellar mass at $z=0$.
Recent work by \cite{Postietal2019} explored galaxies in the SPARC sample \citep{Lellietal2016}, finding that massive spirals lie above the relation for $f_{\rm \star}$ derived from abundance matching, which implies that massive spirals are overly efficient at converting gas to stars (the so-called `failed feedback problem'). A subsequent study by \cite{Marascoetal2020} found that massive spirals in SPARC are more baryon-dominated than spiral galaxies in cosmological simulations such as EAGLE and IllustrisTNG100. Interestingly, the Auriga galaxies follow a similar trend as the high mass spiral galaxies in the SPARC sample, which are denoted by the grey symbols. 

While the $M_{\star}$ -- $M_{\rm h}$ relation tells us about the global ratio of baryons-to-dark matter, the local distribution of baryons-to-dark matter is more relevant for the dynamics of disc galaxies and for the bar instability. We show this in the right panel of Figure \ref{fig:dmbarratio}, where we plot $V_{\star}/V_{\rm tot}$, i.e. the ratio of the stellar component to the total rotation curve, as a function of radius for disc galaxies in Auriga, Illustris and EAGLE (in the mass range $3\times10^{10}<M_{\star}/M_{\odot}<10^{11}$). This reveals how `maximal' or baryon-dominated a galaxy is\footnote{Historically, disc maximality has been calculated at a given radius, such as at 2.2 disc scalelengths.}, which in turn will partially determine whether the disc is able to form a bar that remains fast. 
 We find that the Auriga barred galaxies are overall more baryon-dominated than EAGLE and Illustris at all radii.
In Figure \ref{fig:fstarhighz} we explore this trend also at redshift $z=0.5$ and find that Auriga galaxies are already more baryon-dominated at higher redshifts. 

What causes Auriga to be more baryon-dominated than the EAGLE and Illustris simulations is the complex interplay of the subgrid implementations of the baryonic physics in the simulations. A full exploration of these is well beyond the scope of this letter, but here we outline a few differences that are likely important, and refer the reader to the papers describing the simulations for more details \citep{Schayeetal2015,Vogelsbergeretal2014,Grandetal2017}. In terms of the stellar feedback: the wind model in Auriga has significant differences from its predecessor, Illustris, in both its parametrization and its implementation, e.g. the winds are isotropic in Auriga vs bipolar in Illustris. This might lead to the winds being more effective at higher redshifts in Auriga where galaxies are irregular. On the other hand, EAGLE employs a thermal feedback prescription which is more `bursty', heating particles stochastically to sufficiently high temperatures to avoid catastrophic cooling, which is likely more effective at removing baryons from the central regions of galaxies. In terms of the AGN feedback:the prescription in Auriga is `smoother' than that in Illustris and EAGLE. In Auriga the `bubble' radio-mode feedback provides a more gentle and distributed heating of the circum-galactic medium than in Illustris, in which a smaller number of very energetic bubbles were able to blow out all the gas from the halo; in EAGLE there is only one mode of AGN feedback in which thermal energy is injected stochastically by heating particles around the black hole to very high temperatures.
Furthermore, Auriga includes magnetic fields, in contrast to both EAGLE and Illustris. The differences in these physical modelling assumptions, which combine non-linearly, give rise to discs which are more massive in Auriga.
\section{Conclusions}
\label{sec:concl}

To summarise, we find that bars in the Auriga cosmological simulations are fast, with $\mathcal{R}<1.4$, in agreement with observations, and contrary to the results from previous cosmological simulations.
The considerations of the previous section provide us with a likely explanation for why bars in other cosmological simulations tend to slow down so dramatically by $z=0$: their host galaxies are embedded in dominant dark matter haloes -- both locally and globally -- i.e. they have a higher stellar-to-dark matter ratio, and thus their bars suffer more dynamical friction.
Disc galaxies in the Auriga simulations, on the other hand, are more baryon-dominated, and thus are able to remain fast for over half the age of the Universe. This alleviates the previously reported tension between fast bars and the $\Lambda$CDM cosmological paradigm. However, our findings suggest that in order to reproduce the dynamics of barred galaxies, massive spirals must be baryon-dominated, and should lie above the abundance matching relation. This is consistent with the recently reported findings of \cite{Postietal2019} and \cite{Marascoetal2020}, which highlight the tension between the stellar-to-dark matter ratio of high mass spirals obtained dynamically and via abundance matching. 

Our results imply that, for cosmological simulations of galaxy formation and evolution to reproduce the dynamical properties of observed barred spirals, they need to build up more stellar mass in their discs than that predicted by abundance matching. This suggests that care is needed to not make the stellar and AGN feedback prescriptions overly efficient, as this can lead to lower stellar-to-dark matter ratios in high-mass spirals, which will prevent them from maintaining fast bars. 
As bars are present in a significant fraction ($\geq$ 50\%) of the population of spiral galaxies in the local Universe \citep{Eskridgeetal2000,Menendezetal2007}, this effect is likely to considerably increase the expected scatter around the abundance matching relation, particularly for spirals at the high-mass end, where bars are more numerous \citep{Mastersetal2012,Gavazzietal2015}. Whether these results have repercussions on the expected value of the abundance matching relation or simply on its scatter will depend on the relation followed by the global population of galaxies at this mass range, including early type galaxies. In terms of the implications for lower mass galaxies, there is still considerable debate in the observational community with regards to the frequency of bars in low mass spirals \citep{NairAbraham2010,Erwin2018}, and as such, an exploration of the properties of lower mass barred spirals is timely. 
Our results highlight the importance of taking into account the dynamics of barred spiral galaxies when constraining models of galaxy formation and evolution.
  
\begin{acknowledgements}
We thank the anonymous referee for a constructive report. FF thanks L. Posti, B. Moster and C. Damiani for useful discussions and N. Peschken for providing the halo IDs of barred galaxies in Illustris. FM acknowledges support through the Program ``Rita Levi Montalcini'' of the Italian MUR.
\end{acknowledgements}

%
%

\bibliographystyle{aa}
\bibliography{References}%

\begin{appendix} 

\section{Barred galaxies in the Auriga simulations}
\label{sec:appAu}
The Auriga simulations \citep{Grandetal2017} are a suite of 30 magneto-hydrodynamical cosmological zoom simulations of haloes with masses in the range of $1 \times 10^{12}-2 \times 10^{12}M_{\odot}$ which run from redshift $z=127$ to $z=0$ with cosmological parameters $\Omega_m=0.307$, $\Omega_b=0.048$, and a Hubble constant of $H_0=67.77$ km$ \, \rm s^{-1} \, \rm Mpc^{-1}$ \citep{Planck2014XVI}; unless otherwise stated we use these parameters throughout the paper.
The simulations are performed with the magneto-hydrodynamic code AREPO \citep{Springel2010,Pakmoretal2016}, with a comprehensive galaxy formation model \citep{Vogelsbergeretal2013, Marinaccietal2014a,Grandetal2017} and form star-forming disc galaxies with flat rotation curves that reproduce a range of observed scaling relations such as the Tully-Fisher relation \citep{Grandetal2017} and the size-mass relation of HI gas discs \citep{Marinaccietal2017}. They form structures such as bars and boxy/peanuts which have properties compatible with those of observed bars \citep{BlazquezCaleroetal2020,Fragkoudietal2020}  and mainly consist of so-called pseudo-bulges \citep{Gargiuloetal2019}. For more details we refer the reader to the aforementioned papers and references therein.

In this study our `barred sample' includes all the Auriga galaxies with bar strength $A_2>0.2.$ at $z=0$. This excludes five Auriga galaxies which are undergoing an interaction at $z=0$ (Au1, Au11, Au20, Au25 and Au30). For these galaxies the pattern speed cannot be reliably measured using the Tremaine-Weinberg method, for which the continuity equation must hold. We therefore have 16 Auriga galaxies in our barred sample. Five of these haloes (Au9, Au17, Au18, Au26 and Au28) are re-runs of the original Auriga haloes with higher cadence outputs (i.e. every 10\,Myr), in order to be able to determine the pattern speed from the temporal evolution of the bar and to test our implementation of the Tremaine-Weinberg method (due to the high computational cost we do do not re-run all the Auriga galaxies). For the re-runs, the initial conditions of the haloes and the physics implementations are the same as those of the original Auriga haloes.
We refer the reader to the Appendix where we describe in detail how we obtain the bar strength, length and pattern speed of our simulated galaxies.

\section{Bar strength, length \& corotation radius}
\label{sec:ap1}

\begin{figure}
    \begin{center}
        \includegraphics[width=0.45\textwidth]{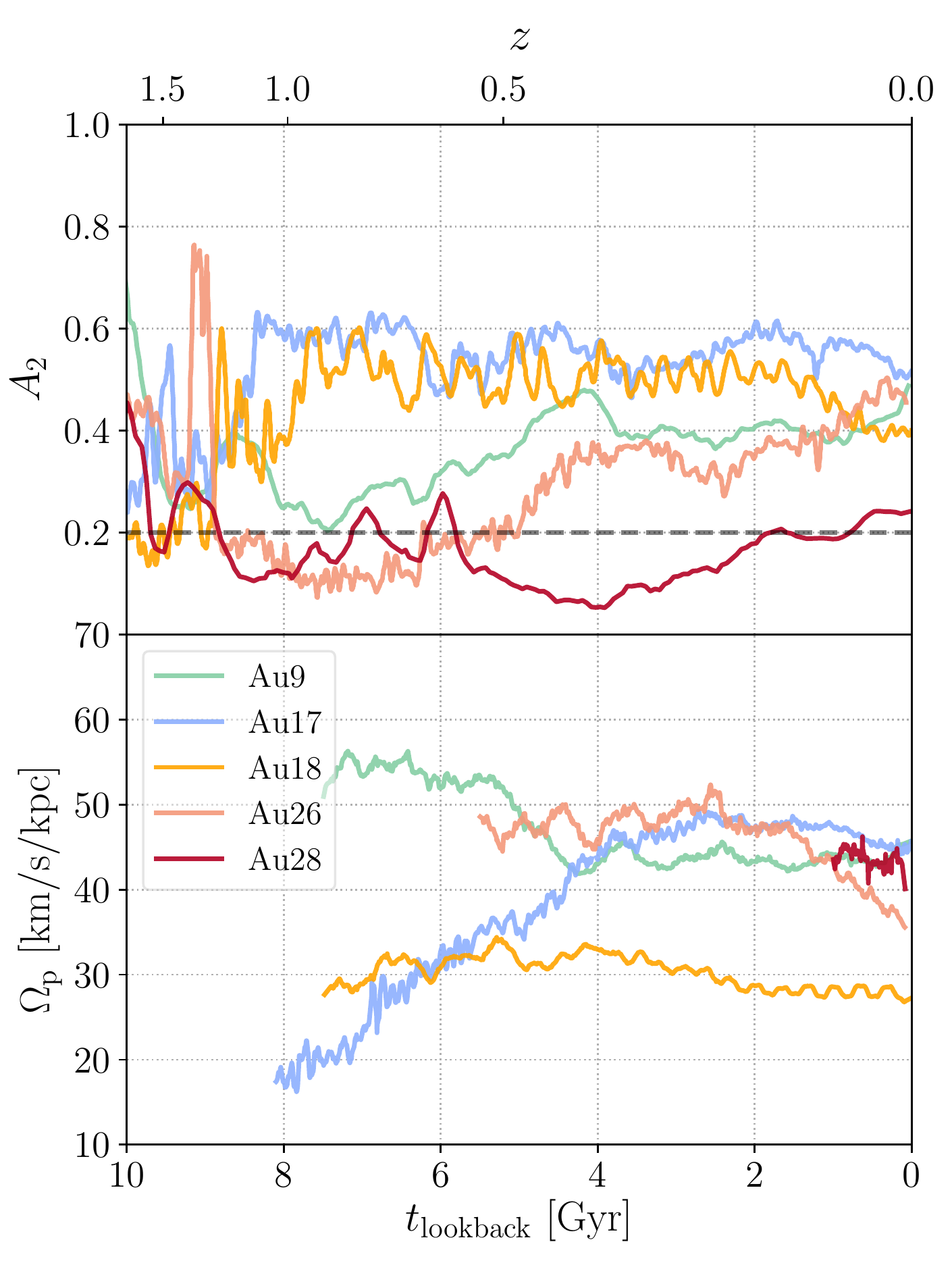}
        \caption{{\textbf{Bar properties as a function of time:}} \emph{Top panel:} Bar strength for the five Auriga galaxies for which we have high cadence outputs as a function of lookback time. The horizontal dashed line indicates $A_2=0.2$ above which we consider the bar to have formed. \emph{Bottom panel:} Bar pattern speed for these galaxies, once the bar is formed, as a function of lookback time.} 
        \label{fig:a2omtime}
    \end{center}
\end{figure}

\paragraph{Bar Strength}
We define the bar strength as the maximum amplitude of the normalised $m=2$ Fourier mode of the surface density. We select the stellar particles in the disc (i.e. with $|z| < 1\,\rm kpc$) in annuli of width 0.5\,kpc to calculate,
\begin{equation}
a_m(R) = \sum^{N}_{i=0}  {\rm m_i} \cos(m\theta_i), \,\,\,\,\, \,\,\,\,\, m=0,1,2,...,
\end{equation}
\begin{equation}
b_m(R) = \sum^{N}_{i=0} {\rm m_i} \sin(m\theta_i),  \,\,\,\,\,\,\,\,\,\, m=0,1,2,...
\end{equation}

\noindent where $\rm m_i$ is the mass of particle $i$, $R$ is the cylindrical radius, $N$ is the total number of particles in that annulus and $\theta$ is the azimuthal angle.
To obtain a single value for the bar strength at each time-step, we take the maximum value of the normalised $m=2$ component,
\begin{equation}
A_2 = \mathrm{max} \left( \frac{\sqrt{ a_2^2(R) + b_2^2(R) } }{a_0(R)} \right)
\end{equation} 
in radius.

We also carry out visual inspections of the snapshots in order to ensure that the $m=2$ mode is due to the bar, and not some other spurious short-lived effect. For example, at high redshifts -- when mergers are frequent -- short-lived $m=2$ modes can appear due to the centre of mass of the merging system not being at the centre of the disc (e.g. in the final stages of a merger). This would therefore lead our algorithm to detect a spurious and short-lived $m=2$ mode, which is however not related to a bar mode in the disc. However we note that such spurious and short-lived spikes in $m=2$ are essentially only found at high redshifts (see e.g. the large peak in $m=2$ for Au26 at $\sim$9\,Gyr in Figure \ref{fig:a2omtime}).

\begin{figure}
    \begin{center}
        \includegraphics[width=0.4\textwidth]{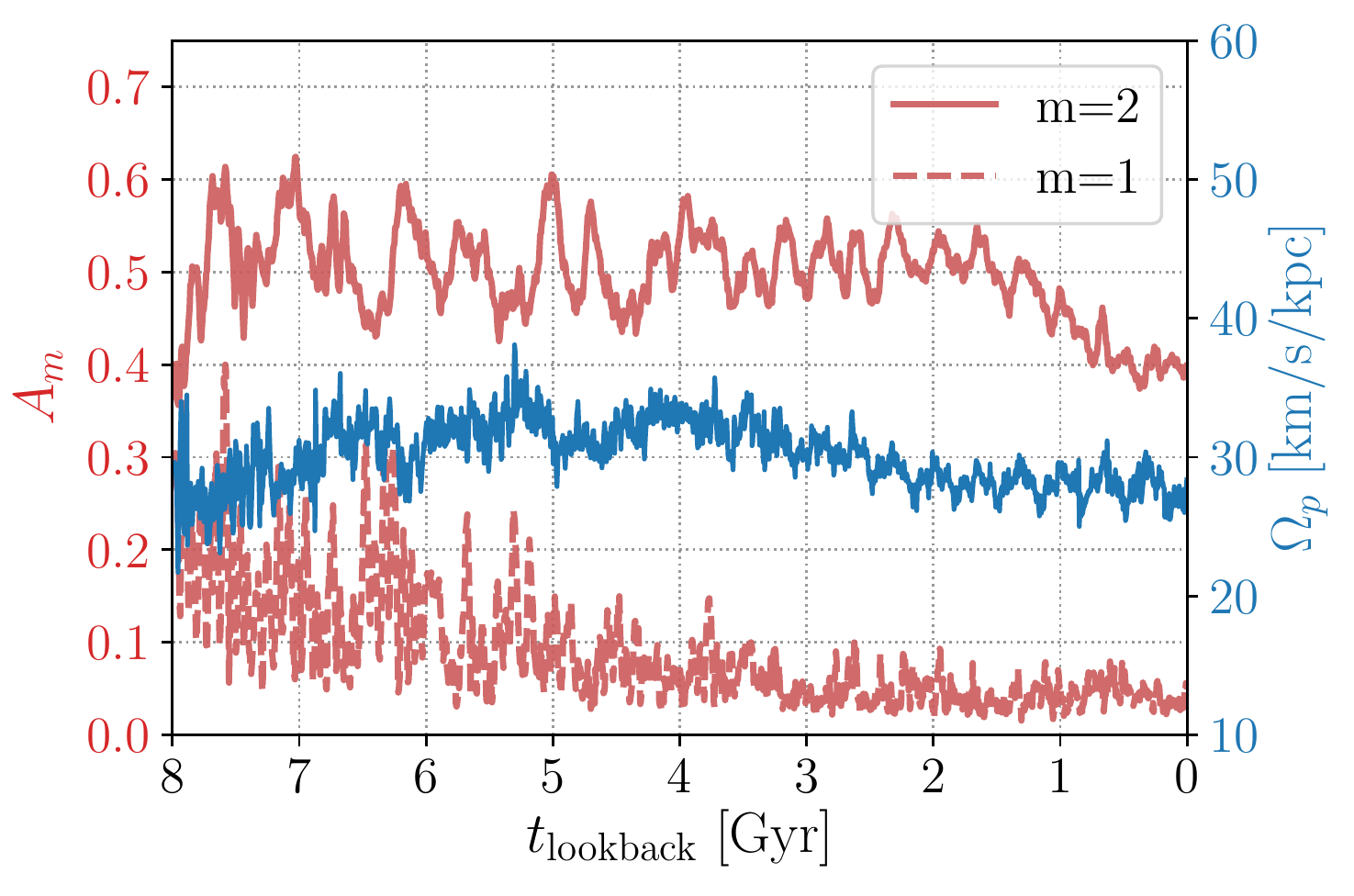}
        \includegraphics[width=0.4\textwidth]{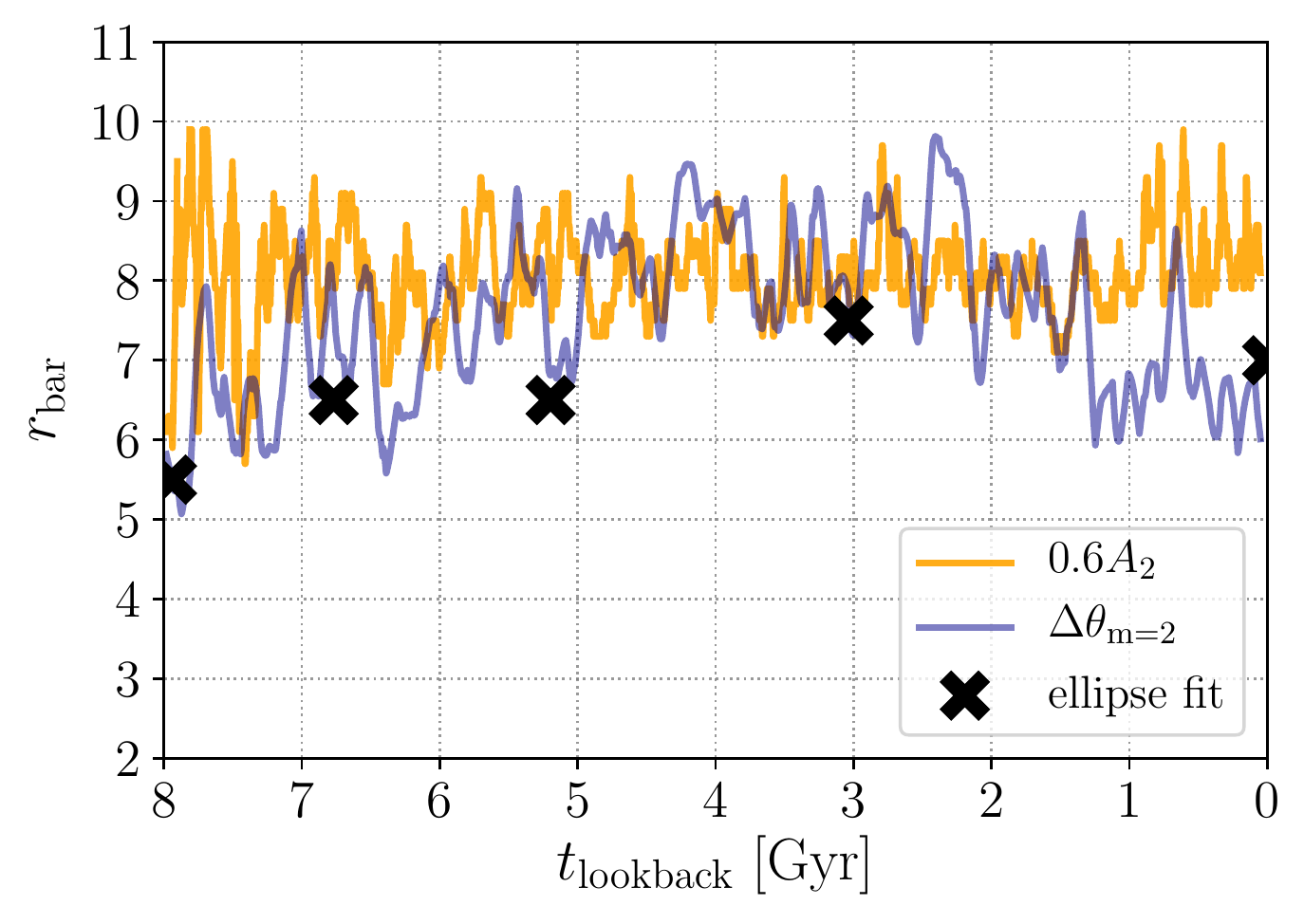}
        \includegraphics[width=0.4\textwidth]{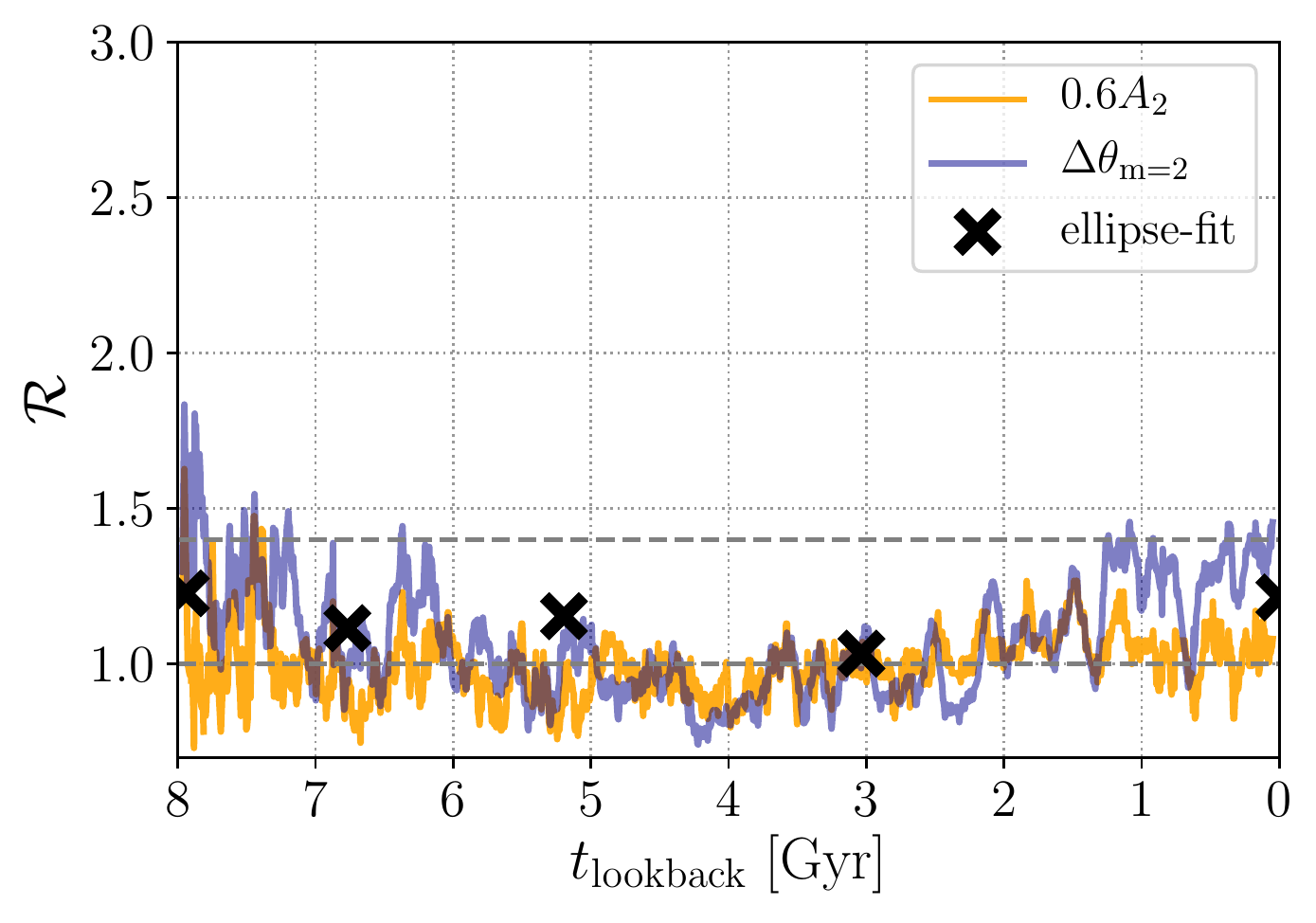}
        \caption{Bar strength, pattern speed and length variation for Au18 with time: \emph{Top panel:} We show the pattern speed as estimated from the phase of the $m=2$ Fourier mode (righ axis -- blue), the $m=2$ Fourier mode (bar strength) of the surface density, as well as the $m=1$ Fourier mode (dashed red line), as a function of lookback time. \emph{Middle panel:} Various measures of bar length as a function of lookback time (see text). \emph{Bottom panel:} Variation of $\mathcal{R}$ as a function of time using different estimates for the bar length. } 
        \label{fig:slp_var}
    \end{center}
\end{figure}

\paragraph{Pattern Speed}
For the five galaxies for which we have high cadence outputs the bar pattern speed is obtained by examining the temporal evolution of the bar phase, i.e. by calculating the $m=2$ phase from the aforementioned Fourier decompositions in each snapshot and as a function of radius as,

\begin{equation}
\theta (R) = 0.5 \arctan \left( \frac{b_2(R)}{a_2(R)}\right).
\end{equation}

The bar pattern speed $\Omega_{\rm p}$ is then calculated as an annular average within the bar radius as, 
\begin{equation}
\Omega_{\rm p} = \frac{\Delta \langle \theta \rangle}{\Delta t}.
\end{equation}

The uncertainty on this value mainly derives from the oscillations in pattern speed at different snapshots (see for example the bottom panel of Figure \ref{fig:a2omtime}), which are typically around $\pm$3-5km/s/kpc. We explored the uncertainty this implies in terms of the corotation radius and found that this translates to $\sim\pm$0.5\,kpc. 

Changes in the pattern speed as a function of time can occur due to coupling with spiral arms or odd modes such as $m=1$. To check whether $m=1$ can have a significant effect on our estimates of the pattern speed, in the top panel of Figure \ref{fig:slp_var} we show the variation of the $m=2$ and $m=1$ modes as well as the pattern speed as a function of lookback time. We find that at higher redshifts, when gas accretion and minor mergers are more frequent, $m=1$ is non-negligible, and this induces some noise in the pattern speed of the bar. However, by low redshifts the contribution from $m=1$ is negligible, of the order of a few percent, and as such does not seem to affect our estimates of the bar pattern speed -- especially as we exclude from our sample the galaxies which are undergoing interactions at $z=0$.

In Figure \ref{fig:a2omtime} we show the evolution of the bar strength and pattern speed as a function of lookback time for five of the Auriga galaxies for which we have high cadence outputs. In the top panel of the Figure we show the evolution of the bar strength.
Bars form at a range of lookback times in our simulations, with some forming at $z>1$ (e.g. Au17 and Au18), or as late as 1\,Gyr ago (Au28). 
Bars in Auriga are long-lived structures, i.e. once the bar has formed it does not dissolve. 
In the bottom panel of Figure \ref{fig:a2omtime} we show the bar pattern speed, $\Omega_{\rm p}$, for these five barred galaxies as a function of lookback time. Most of the bars have pattern speeds which remain roughly constant or are slightly decreasing (e.g. Au26 after $t_{\rm lookback}\sim2\,\rm Gyr$). There is also the curious case of Au17, in which the bar pattern speed increases between formation time and until $t_{\rm lookback}\sim3\,\rm Gyr$ due to a resonant interaction with another nearby massive system (a more detailed exploration of this system will be presented elsewhere). The galaxy with the largest decrease in $\Omega_{\rm p}$, Au26, shows a prototypical example of bar growth in an isolated disc, i.e. while the galaxy evolves in isolation the bar grows gradually stronger over time, while $\Omega_{\rm p}$ decreases. In a number of the other cases shown here, bar formation is triggered after a significant merger, with the disc losing a large fraction of its angular momentum by torques induced during the merger \citep{Fragkoudietal2020}, with the bar subsequently not growing much stronger or slower. 

For the rest of the Auriga barred sample for which we do not have high cadence outputs we employ the commonly-used Tremaine-Weinberg method \citep{TremaineWeinberg1984}, which uses the continuity equation to estimate the bar pattern speed from a single snapshot at $z=0$. We tested our method on the five simulations for which we have high cadence outputs -- and therefore a reliable estimate for the bar pattern speed -- and found that we can typically recover the pattern speed to within $\pm$5km/s/kpc. This translates to an error on the corotation radius of $\sim\pm$1\,kpc. These are the uncertainties employed in Figure \ref{fig:Omprbar}.

\begin{figure*}
\centering
\includegraphics[height=0.24\textwidth]{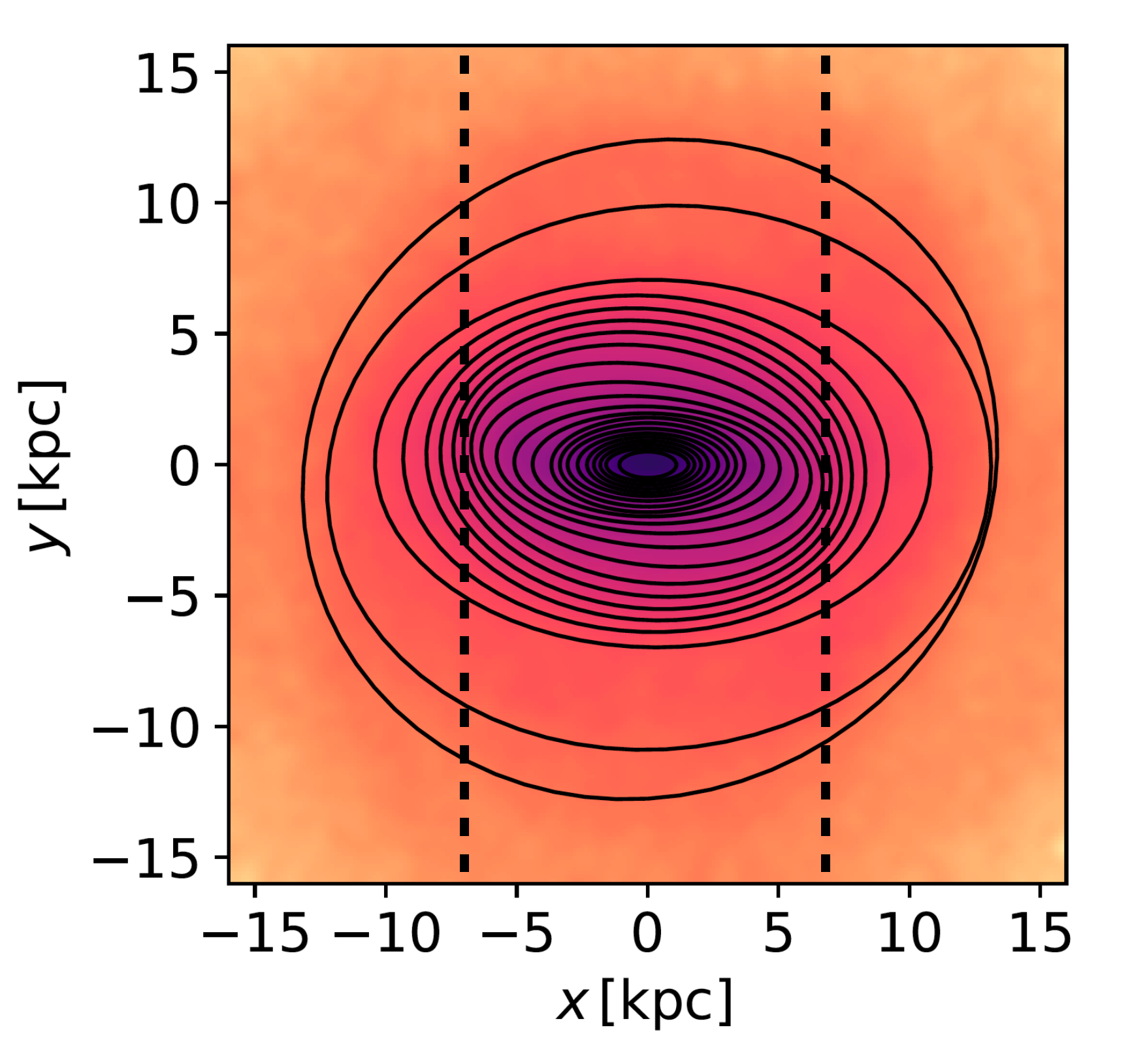}
\includegraphics[height=0.24\textwidth]{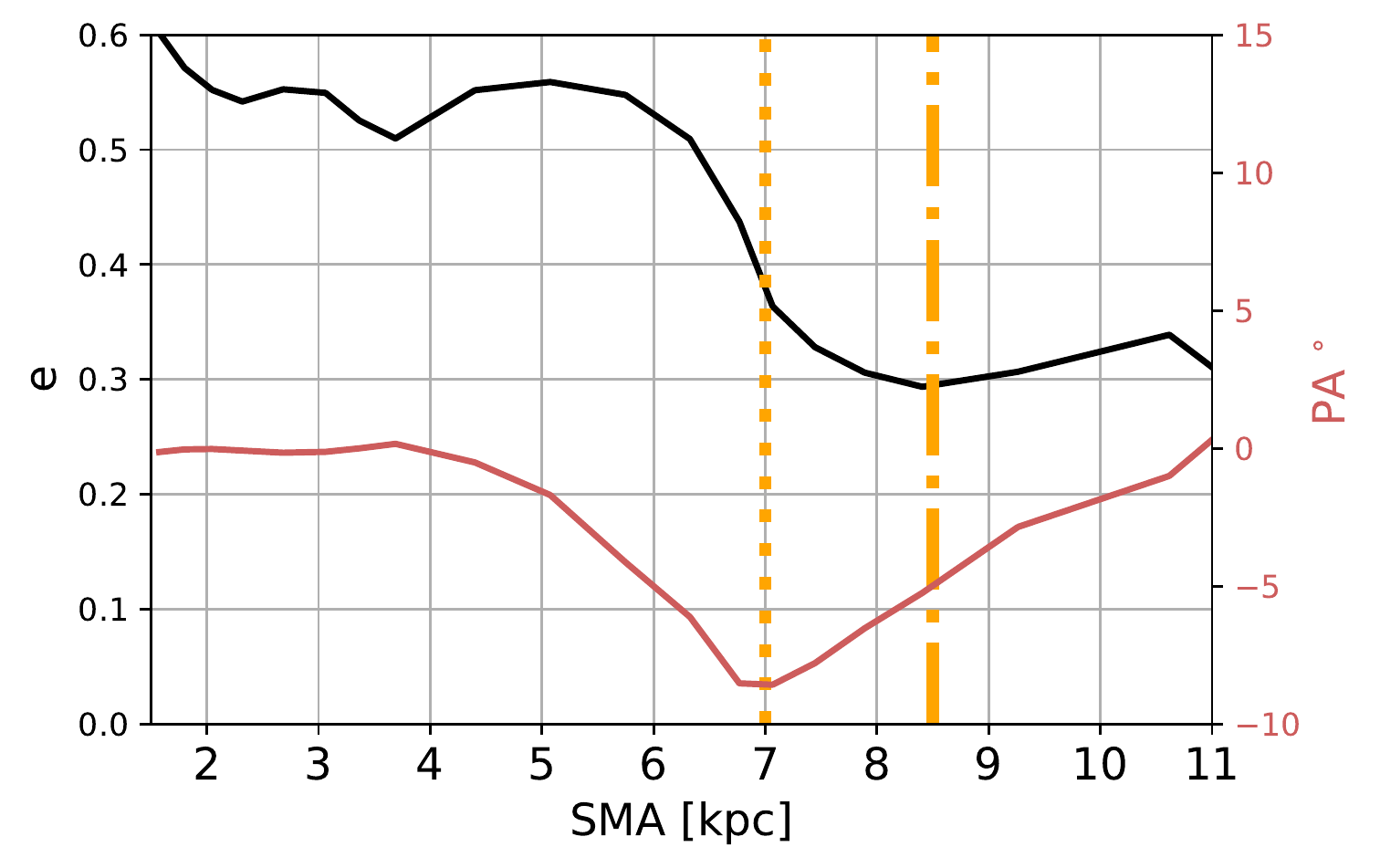}
\includegraphics[height=0.24\textwidth]{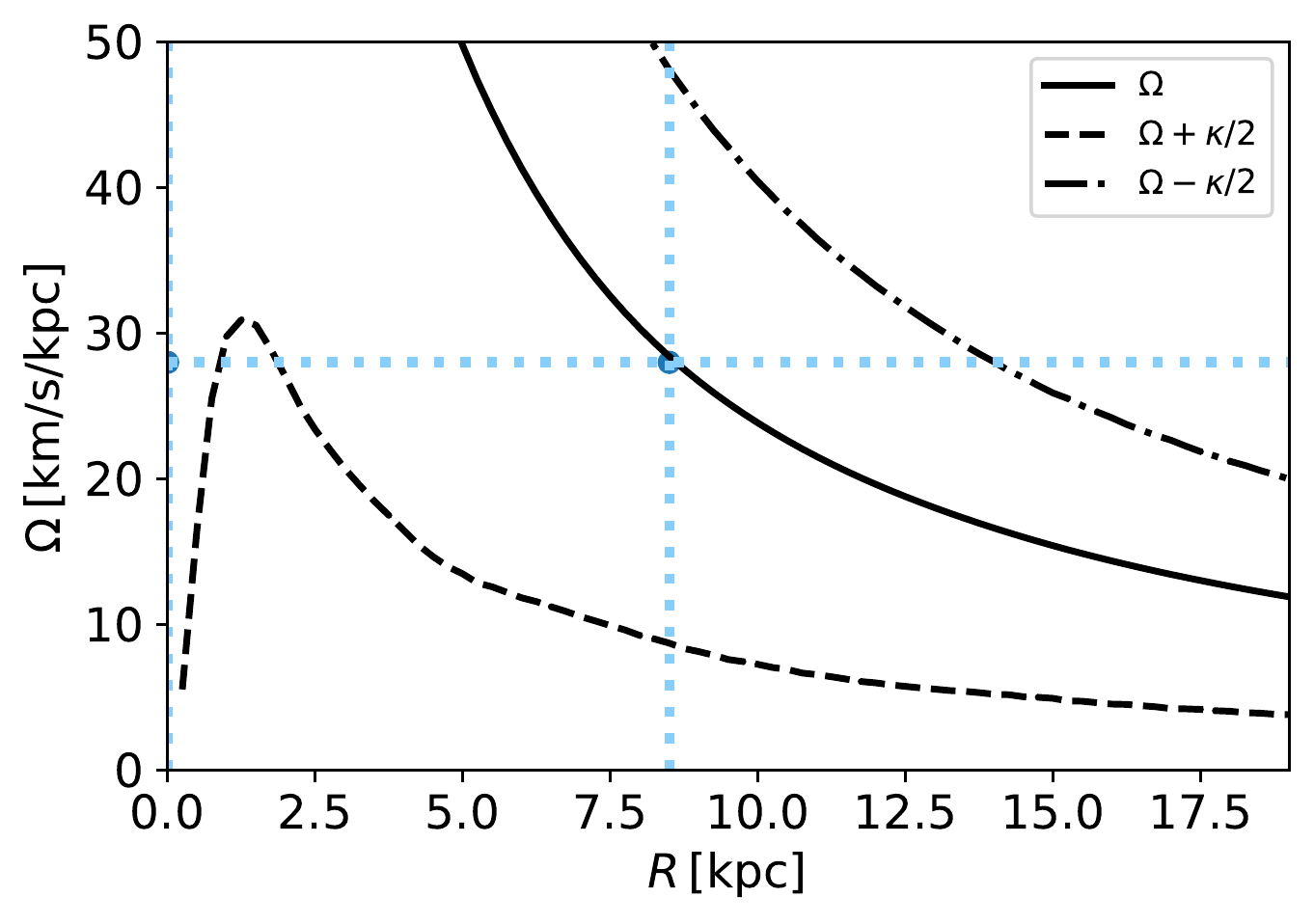}
\caption{{\bf{Estimating the bar length and corotation radius:}} \emph{Left panel:} The bar length is estimated by fitting ellipses to iso-density contours. The vertical dashed lines denote the estimated bar length. \emph{Middle panel:} Ellipticity and position angle as a function of semi-major axis of the ellipses. The bar length is the minimum of two values: the radius of the first minimum after the ellipticity maximum (dot-dashed orange line) or the radius at which there is the first extremum in position angle outside the bar (dashed line). \emph{Right panel:} The bar corotation radius (marked by the vertical dashed blue line) is the radius at which stars in the disc have the same angular frequency (solid black line) as the bar pattern speed (horizontal dashed blue line). The dashed and dot-dashed lines denote the $\Omega-\kappa/2$ and $\Omega+\kappa/2$ curves for the galaxy.} 
\label{fig:calcbarlength}
\end{figure*}

\paragraph{Bar Length}
There is no standard way for estimating the bar length and various methods have been used in observations and simulations -- e.g. visual estimation, isophote ellipse fitting, measurements based on Fourier analysis etc. -- each with its own advantages and disadvantages (for a discussion of the various methods see \citealt{Erwin2005,Petersenetal2019,Hilmietal2020}). Here we calculate the bar length using a method commonly used in observations (such as the observations presented in Figure \ref{fig:Omprbar}), i.e. we fit ellipses to isodensity contours of face-on images of the galaxies. 

We follow a prescription similar to the one used in \cite{Erwin2005}, in which the bar length is defined as the minimum of two ellipse fitting measures: the radius of the first minimum after the maximum of ellipticity, or the radius at which the position angle has an extremum outside the bar. An example of this is shown in the left and middle panels of Figure \ref{fig:calcbarlength}. If there is no extremum in position angle then we use instead the radius at which the position angle changes by 10 degrees. For this technique, we do not implement an automated method, but inspect the outputs on a snapshot by snapshot basis. Therefore this method is not efficient at obtaining bar lengths for a large number of snapshots, but it does give a rather accurate bar length estimate. Other techniques which are more conducive to being automatised exist, such as defining the bar length as the radius at which $A_2$ drops below a certain percentage of its maximum value, or the radius at which there is a certain change in the phase $\theta$ of the $m=2$ component. These methods however can lead to an over- or under-estimation of the bar length, due to the complex interplay between various modes in the disc; for example, when the spiral arms and bar align this can lead to an overestimate of the bar length (see e.g. \citealt{Hilmietal2020} for a detailed exploration of various estimates of the bar length and the effects of bar-spiral coupling on these estimates), or an underestimation of the bar length can occur when a secondary mode (such as an inner bar) forms with a different pattern speed.

To explore whether the value of $\mathcal{R}$ we obtain for the Auriga galaxies in this work would be significantly altered if we used a different measure of the bar length, we explore, in the middle and bottom panels of Figure \ref{fig:slp_var}, different bar length estimates. We employ two commonly-used methods from the literature, i.e. when the $m=2$ Fourier mode drops below 60\% of the maximum (orange), and when the phase of the $m=2$ Fourier mode changes by more than $\Delta \theta_{m=2}$=7.5 degrees (blue). Employing different values for the $A_2$ threshold or $\Delta \theta_{m=2}$ does not qualitatively affect these results. The estimates we use based on the ellipse fitting method (and which is used in the main part of the paper) are shown with black crosses. In the middle panel of Figure \ref{fig:slp_var} we see that our ellipse-fitting method for estimating the bar length coincides with the lower values obtained from the other two methods, in agreement with what is estimated as closer to the `true' bar length in \cite{Hilmietal2020}. The bar length we use in the main part of the paper therefore gives values on the low side of that obtained from the fluctuations in bar length due to bar-spiral coupling. This in turn leads to higher values of $\mathcal{R}$, as is shown in the bottom panel of Figure \ref{fig:slp_var}, i.e. for most of the temporal range our method gives consistent or even slightly higher values of $\mathcal{R}$ than the other two methods. Thus, we see that using another method to estimate the bar length would not change the results of our study by any significant amount, i.e. we would not find values of $\mathcal{R}>1.4$, thus changing our conclusion on whether Auriga bars are fast or slow. We note that at $z\sim0$ the algorithm which obtains the bar length from the difference in the $m=2$ phase ($\Delta \theta_{m=2}$) gives shorter bar estimates than the one we obtain from our ellipse-fitting method and the maximum of $A_2$ method. This is due to the growth of an `inner bar'-like structure at late times in the galaxy, which has a different pattern speed than the `main' bar of the galaxy. This leads to abrupt phase changes at low radii. However, upon visual inspection it is clear that the primary bar is longer.

\paragraph{Corotation Radius}
To obtain the corotation radius of Auriga barred galaxies we first need to calculate the circular velocity of the galaxy $V_{\rm c}^2(r) = r \frac{\partial \Phi}{\partial r}$.  
We make the simplifying assumption of spherical symmetry, which reduces the circular velocity to $V_{\rm c} = \sqrt{GM(r)/r}$ where $M(r)$ is the enclosed mass within a sphere of a radius $r$. This simplification will tend to under-predict the true circular velocity, which will be higher for flatter systems. We tested this by assuming a Miyamoto-Nagai disc \citep{BT2008} with realistic flattening of the disc, for which we can calculate $V_{\rm c}$ analytically. We then calculated the approximate value of $V_{\rm c}$ using the spherical symmetry assumption. We find that the error in $V_{\rm c}$ is of the order of $\sim5-10\%$ depending on the flattening of the disc -- with the error increasing for thinner discs. For galaxies in our sample which also have a spherical component due the dark matter halo this error will be smaller, roughly of the order of 5\%, which is acceptable for the purposes of this study. The angular frequency of stars is then obtained according to $\Omega = V_{\rm c}/r$ and the epicyclic frequency is calculated as 
$\kappa = \sqrt{R\frac{d\Omega^2}{d R} + 4\Omega^2}$ \citep{BT2008}. The corotation radius is the radius at which stars on nearly circular orbits in the disc have the same angular frequency as the bar. We therefore calculate this as the intersection between the $\Omega$ curve with the bar pattern speed $\Omega_{\rm p}$ (right panel of Figure \ref{fig:calcbarlength}).

We note that for the $\mathcal{R}$ values of barred galaxies in EAGLE and Illustris shown in the bottom panel of Figure \ref{fig:Omprbar}, we use the values reported in \cite{Algorryetal2017} and \cite{PeschkenLokas2019} respectively. For the Illustris $\mathcal{R}$, we divide the value reported in \cite{PeschkenLokas2019} by 2, since -- as stated by the authors -- they used a proxy for the bar length which underestimates the true bar length by a factor of $\sim$2.

\section{Estimating the galaxy flattening}
\label{sec:dth}

Galaxies with thinner disks (i.e. with a smaller scale-height to scale length ratio) have been found to host fast bars, compared to models with thicker discs which appear to be slower \citep{MisiriotisAthanassoula2000,Klypinetal2009}. We therefore explore whether disc thickness could be the main culprit for making bars in Auriga faster as compared to Illustris or EAGLE. We calculate the thickness of the discs in Auriga, and compare to publicly available values obtained for Illustris and EAGLE (see \citealt{Geneletal2015} and \citealt{Thobetal2019}). For Illustris and EAGLE we take only discs galaxies in the mass range $3\times 10^{10}<M_{\star}/M_{\odot}<10^{11}$, i.e. the mass range covered by the Auriga galaxies.

To calculate the disc thickness in Auriga we employ the same method as that in Illustris and EAGLE, i.e. we model the galaxies as three dimensional ellipsoids, and calculate their reduced inertial mass tensor,

\begin{equation}
    I_{ij}= \frac{1}{M}\sum_n m_n \frac{r_{n,i} r_{n,j}}{r^2_{n}}
\end{equation}

\noindent where $i,j = \{0,1,2\}$ denote the three principal axes, $r_{n,i}$ is the $i$th component of the coordinate vector of particle $n$, $m_n$ is the particle’s mass and $M$ is the total mass of the galaxy within the radius we consider. To calculate the reduced mass tensor we use only stellar particles inside 2$\times r_{50}$, where $r_{50}$ is the half-mass radius of the galaxy.
The axis lengths of the galaxy are then defined as the square root of the eigenvalues $\lambda_i$ of this matrix. We then compare the flatness of the discs in the different simulations, defined as $\epsilon = 1 - c/a$, where $c$ denotes the smallest axis, and $a$ the largest axis of the ellipsoid. The results of this comparison can be seen in Figure \ref{fig:restest}b.

\section{The abundance matching relation, $M_{\rm \star}$ and $M_{\rm h}$ and maximality}
\label{sec:ap3}

\begin{figure}
\centering
\includegraphics[width=0.49\textwidth]{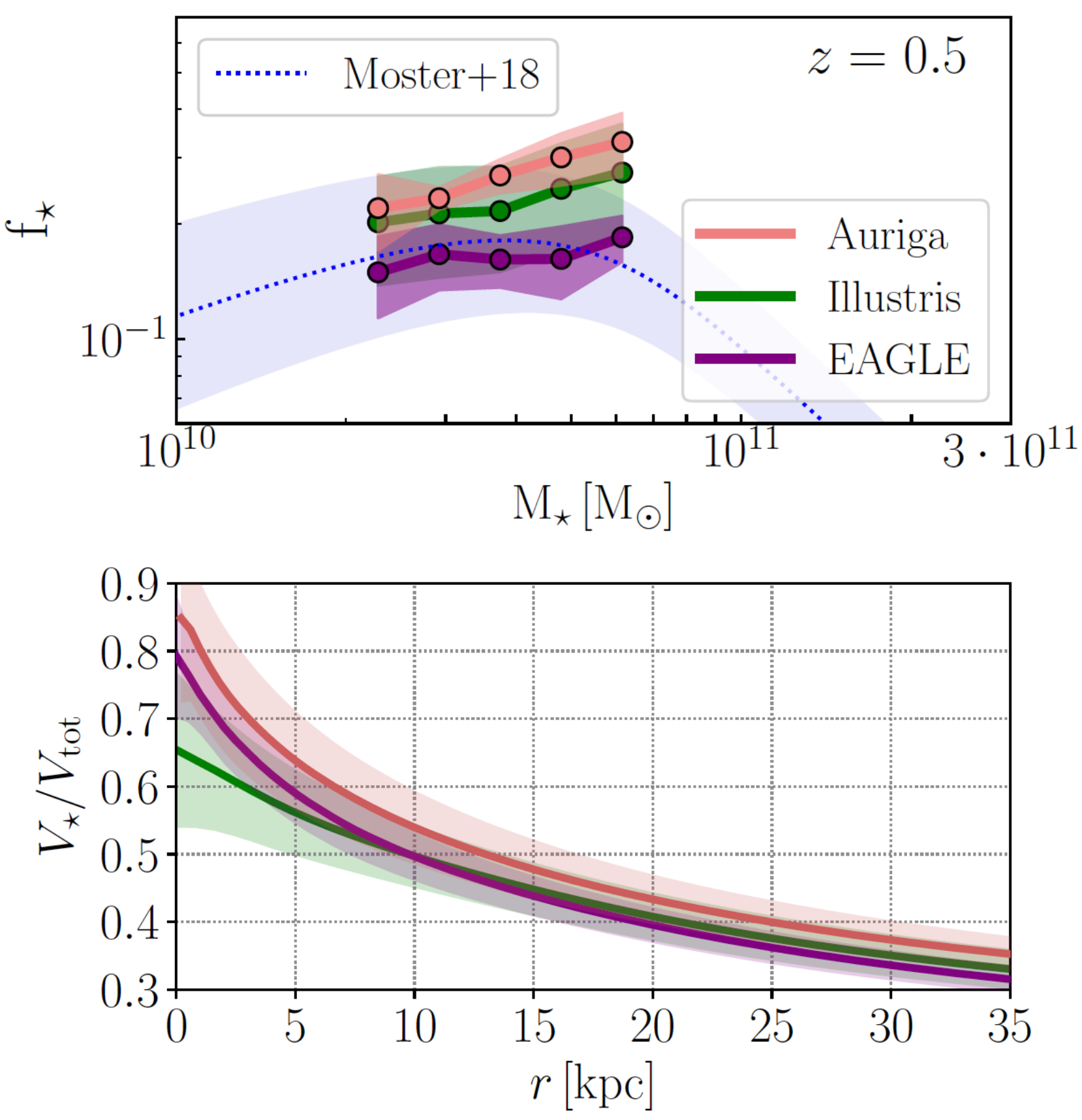}
\caption{{\textbf{Stellar to dark matter ratios at higher redshifts:}} \emph{Top:} $f_{\star}$ vs. $M_{\star}$ for Auriga, EAGLE and Illustris disc galaxies at $z=0.5$. The dotted blue line denotes the relation determined using the Moster et al. (2018) abundance matching relation and the shaded region denotes the 1$\sigma$ scatter around the relation. \emph{Bottom} $V_{\star}/V_{\rm tot}$ for the galaxies shown in the top panel. Auriga is locally more baryon-dominated than EAGLE and Illustris already at higher redshifts.}
\label{fig:fstarhighz}
\end{figure}

In Figures \ref{fig:dmbarratio} and \ref{fig:fstarhighz} we use the abundance matching relation from Moster et al. (2018), employing the values for ``All centrals'' from their Table 8 and using their derived relation for the scatter in equation 25 \citep{Mosteretal2018}. The values of $M_{\rm \star}$ and $M_{\rm h}$ for Auriga, Illustris and EAGLE used in the aforementioned figures are obtained as follows:
For Auriga, we calculate $M_{\rm \star}$ by summing the mass of stellar particles inside 10\% of the virial radius of the halo. $M_{\rm h}$ is the total mass of dark matter particles within the virial radius of the halo. 
For Illustris, we extract the $M_{\rm \star}$ within 20\,kpc and the total dark matter mass inside the virial radius $M_{\rm h}$ for disc galaxies using the publicly available Illustris data \citep{Nelsonetal2015}. We use the definition of disc galaxies employed by \cite{PeschkenLokas2019}, i.e. selecting galaxies which have more than 20\% of their stellar mass with circularity parameter above 0.7 and with flattened distributions, i.e. which have a flatness ratio (defined by the ratio eigenvalues of the stellar mass tensor) $<0.7$. For the values extracted for the barred Illustris galaxies, we take as barred Illustris galaxies those studied in \cite{PeschkenLokas2019}, in their "total bar" sample (N. Peschken kindly provided us with these galaxy IDs) where barred galaxies are defined as those disc galaxies with $A_2>0.15$ with some additional criteria (see Section 2.2 of their paper for more details).
For EAGLE, we use the publicly available data from \cite{McAlpineetal2016} and where relevant the subsequent particle data release \citep{EAGLEpublic2017} to obtain $M_{\star}$ (stellar mass within 20\,kpc) and $M_{\rm h}$ (dark matter mass within the virial radius) values for disc galaxies in EAGLE, which we define as having stellar velocity rotation-to-dispersion ratios (\emph{RotToDispRatio}), and the stellar disc-to-total ratio from counter rotation (\emph{DiscToTotal}) larger than 1.7 and 0.7, respectively. For the barred galaxies, we use the IDs of the galaxies identified as barred in \cite{Algorryetal2017}.  

To explore the local baryon dominance of barred galaxies in Figure \ref{fig:dmbarratio}, we follow the definition of disc maximality used in previous works, i.e. we employ the ratio of the circular velocity due to the stellar component as compared to the total circular velocity. We obtain the stellar and total circular velocities as described in the previous section.

\end{appendix}


\end{document}